\documentclass[reprint, notitlepage,aip,onecolumn,amsmath,amssymb,10pt]{revtex4-2}

\usepackage{amsmath,amsfonts,amssymb}
\usepackage{graphicx}
\usepackage{color,soul}
\usepackage{xcolor}
\usepackage{bm}

\begin{document}

\title{Experimental Realization of Schumacher’s Information Geometric Bell Inequality}

\author{Tahereh Rezaei }
\affiliation{Department of Physics, Florida Atlantic University, Boca Raton, FL, 33431}

\author{Shahabeddin~M.~Aslmarand}
\affiliation{Department of Physics, Florida Atlantic University, Boca Raton, FL, 33431}

\author{Robert Snyder}
\affiliation{Department of Physics, Florida Atlantic University, Boca Raton, FL, 33431}
\affiliation{L3Harris, 10001 Jack Finney Blvd, CBN018, Greenville, TX 75402, USA.}

\author{Behzad Khajavi}
\affiliation{Dept of Biomedical Engineering, University of Houston, Houston, Texas, 77204}

\author{Paul~M.~Alsing} 
\affiliation{Air Force Research Laboratory, Information Directorate,Rome, NY 13441}

\author{Michael~Fanto}
\affiliation{Air Force Research Laboratory, Information Directorate,Rome, NY 13441}

\author{Doyeol (David) Ahn}
\affiliation{Center Quantum Information Processing, Department of Electrical and Computer Engineering, University of Seoul, Seoul 130-743, Republic of Korea}
\affiliation{Department of Physics, Florida Atlantic University, Boca Raton, FL, 33431}

\author{Warner A. Miller}
\affiliation{Department of Physics, Florida Atlantic University, Boca Raton, FL, 33431}

\begin{abstract}
{\bf Abstract}\\Quantum mechanics can produce correlations that are stronger than classically allowed. This stronger--than--classical correlation is the ``fuel'' for quantum computing.  In 1991 Schumacher forwarded a beautiful geometric approach, analogous to the well-known result of Bell, to capture non-classicality of this correlation for a singlet state. He used well-established information distance defined on an ensemble of identically--prepared states. He calculated that for certain detector settings used to measure the entangled state, the resulting geometry violated a triangle inequality --- a violation that is not possible classically.  This provided a novel information--based geometric Bell inequality in terms of a ``covariance distance.''  Here we experimentally-reproduce his construction and demonstrate a definitive violation for a Bell state of two photons based on the usual spontaneous parametric down-conversion in a paired BBO crystal. The state we produced had a visibility of $V_{ad}=0.970$. We discuss generalizations to higher dimensional multipartite quantum states.
\end{abstract}

{
\let\clearpage\relax
\maketitle
}


\section{\uppercase{Quantum Entanglement and the Geometry of Entanglement}}
\label{sec:intro} 

The pithy phrases, ``It--from--Bit'' and ``Information is Physical," of John Archibald Wheeler and Rolf Landauer; respectively, beautifully capture the central role that information plays in our physical laws \cite{Wheeler:1990,Landauer:1991, Zurek:1989}. The quantum phenomenon is strange. It has no localization in space or time. It is timeless in the sense that one can swap the temporal order of conditional measurements with unitary gates in quantum circuits. It is strange because it yields a pure yes, no character. Quantum phenomenon is more deeply dyed with information--theoretic character than anything else we know in physics. This leads us to ask what lies behind and beneath the elementary quantum phenomenon, and also what role it has as a building block in constructing the particles, fields, and geometry of physics \cite{Miller:1983}. This information--based quantum physics is emerging today as a pathway to unify quantum mechanics and general relativity through entanglement \cite{Maldacena:1998,Raamsdonk:2010,Qi:2018,Han:2019}. 

With the emergence of quantum technologies with higher and higher complexities, it may be prudent to examine quantum networks and algorithms from an information perspective. This is our motivation for this research, to begin with, the first step toward a series of information geometry probes into quantum networks, i.e. an entropic--based correlation measure that has metric properties. Since quantum correlation and entanglement is the key resource for quantum information processing, it would seem prudent to explore measures of this resource from an information perspective. Many researchers have explored this direction of research; however, we are unaware of a single entanglement measure that has shown scalability and satisfied necessary general invariance properties. 

Often we find that geometry is a good guide to solve such problems. We concern ourselves here with an experimental realization of one such approach that may help in our future development. We examine the triangle inequality introduced by Benjamin Schumacher that is based on measurements of a Bell state \cite{Schumacher:1991}. His approach was an innovative application of quantum information geometry that highlighted quantum entanglement between two qu$b$its. It is the goal of this manuscript to experimentally reproduce Schumacher's inequality for a near maximally--entangled quantum state approximating the Bell state $|\Phi^+\rangle=(|00\rangle+|11\rangle)/\sqrt{2}$. We produce our state using two photons from a spontaneous parametric downconversion (SPDC) by a paired set of $\beta$--barium borate (BBO) crystals. We recently proposed a generalization of Schumacher's construction from bipartite to multipartite states. In particular, we introduced a geometric-based measure of {\em quantum reactivity} that is a ratio of surface area to volume \cite{Aslmarand:2019,Miller:2018,Aslmarand:2019b, Miller:2019}. Schumacher's original construction for the Bell state was based on a quantum information distance measure of Rolkin and Rajski and later implemented in quantum mechanics by Zurek and Bennett et al. \cite{Rokhlin:1967,Rajski:1961,Zurek:1989,Bennett:1998}. 

The goal of this manuscript is an experimental measurement of Schumacher's triangle inequality. In Sec.~\ref{sec:S} we briefly review Schumacher's 1991 triangle inequality. In the following section, Sec.~\ref{sec:exp} we describe our optical bench setup based on SPDC, and in Sec.~\ref{sec:qsc} measure the quality of the approximate Bell state that our apparatus generates and provide a complete quantum state tomography (QST). The fidelity of our states are approximately $91\%$ and their visibility are $V_{ad}=0.970$. We then show that our results experimentally verify Schumacher's inequality in Sec.~\ref{sec:results}. Finally, in Sec.~\ref{sec:O} we discuss and summarize our results.

\section{\uppercase{Quantum Correlations \& Schumacher's Bipartite Quantum Information Geometry}}
\label{sec:S}

Quantum correlations can be stronger than classical correlations and is based upon the principle of superposition and quantum entanglement. Nature provides us with many potential representations (observables) for any given quantum state, and these can be intrinsic or extrinsic properties or both, e.g. the photons polarization or orbital angular momentum; respectively. For any observable, the principle of complementarity presents us with a binary choice to either measure this quantity or its dual conjugate observable, e.g position or momentum.  It has been shown than one can use these correlations as resource to perform a variety of quantum processes such as quantum key distribution \cite{PhysRevLett.67.661}, quantum teleportation \cite{Bennett:1993} or the improvement of classical protocols such as the reduction of classical communication complexity\cite{Buhrman}, quantum metrology and discrimination\cite{Giovannetti, Bae},  remote state preparation,  quantum locking of classical correlations and quantum illumination \cite{Gerardo}.  

Entanglement is key source for quantum correlations and is considered the most non-classical manifestation of quantum mechanics.  Entanglement has  many strange realizations i.e. Einstein's "spooky action at a distance."  The knowledge of the whole system does not include the best possible understanding of its parts, more precisely, the entropy of the whole quantum state may be less than the sum of its parts.  The entangled quantum system contains correlations that are incompatible with assumptions of classical physics \cite{horodecki, Kurzynski}. These qualities resulted in the famous EPR paper and the idea of an alternative hidden-variable theory of Bohm that has been ruled out by violation of Bell's inequality and its experimental confirmations \cite{EPR, Bell, Aspect, Hensen}. Such experiments showed that idea of local hidden variables, and therefore an incompleteness of quantum mechanics, do not match with experimental results, and this work has been nicely reviewed by Genovese \cite{Genovese}.  Following this broad spectrum of work, scientists were able to solidify the validity of Bell's inequality by introducing loophole-free tests of Bell's inequality \cite{Giustina, Shalm}.

The first step toward utilizing quantum correlations is being able to detect and quantify it. This problem can be approached from two different points of view. One is that given a density matrix, how one can detect which parts of the system are entangled, quantum discord \cite{Zurek}, concurrence \cite{Wootters}, Squashed entanglement \cite{Matthias} are all trying to answer this question. The second approach is that given measurement results how one can detect which parts of the system are entangled Bell and CHSH inequality \cite{CHSH} fall into this group. Both of these questions are important for the field of quantum information, and there is a lot of work done on both subjects. However, in this paper, we are interested in the second question, and we will offer the first experimental proof of a geometrical interpretation of quantum correlation introduced by Schumacher \cite{Schumacher:1991}.

Schumacher used an extension of the Shannon-based information distance, ${\cal D}_{AB} $ defined by  Rokhlin\cite{Rokhlin:1967} and Rajski\cite{Rajski:1961} to quantify the correlation in $\rho_{AB}$, 
\begin{equation}\label{eq:length}
\left(
\begin{array}{c}
Length\ of\\
Edge\ AB
\end{array}
\right) =
{\cal D}_{AB} := H_{A|B}+H_{B|A} = 2 H_{AB} -H_A-H_B,
\end{equation}
where the entropy $H_A$, joint entropy $H_{AB}$, and conditional entropy $H_{A|B}$ are defined as
\begin{align}
\label{eq:ha}
H_A &= -\sum_i p_{a_i} \log{p_{a_i}},\\
\label{eq:hb}
H_{AB} &= -\sum_{i,j} p_{a_ib_j} \log{p_{a_ib_j}},\ \hbox{and}\\
\label{eq:hab}
H_{A|B} &= -\sum_{i,j} p_{{b_j}} \log{p_{a_i|b_j}},
\end{align}
and, without loss of generalization, we will use logarithms base 2 for our numerical calculations. 

Whereas probability measures uncertainty about the occurrence of a single event, entropy provides a measure of the uncertainty of a collection of events. The entropy is the largest when our uncertainty of the value of the random variable is complete (e.g. uniform distribution of probabilities), and the entropy is zero if the random variable always takes on the same certain value,
\begin{equation}
0 \le H_{A} \le1.
\end{equation}
In this sense, entropy is a measure of our ignorance.

The information distance defined in Eq.~\ref{eq:length} is a metric. 
\begin{enumerate}
\item It is constructed to be symmetric, ${\cal D}_{AB}={\cal D}_{BA}$.
\item It obeys the triangle inequality, ${\cal D}_{AB} \le {\cal D}_{Ac}+{\cal D}_{CB}$. 
\item It is nonnegative, ${\cal D}_{AB}\le 0$, and equal to $0$ when  $A$``=''$B$, i.e. when maximally correlated.
\end{enumerate}
Furthermore, if $A$ and $B$ are uncorrelated to each other then, 
\begin{equation}
\label{eq:id} 
{\cal D}_{AB} = 2\left(H_A+H_B\right) - H_A-H_B=H_A+H_B, 
\end{equation}
and ${\cal D}_{AB}$ is bounded,
\begin{equation}
0 \le {\cal D}_{AB} \le H_A+H_B \le 2.
\end{equation}

\begin{figure} [ht]
   \begin{center}
   \begin{tabular}{c} 
   \includegraphics[height=2.0in]{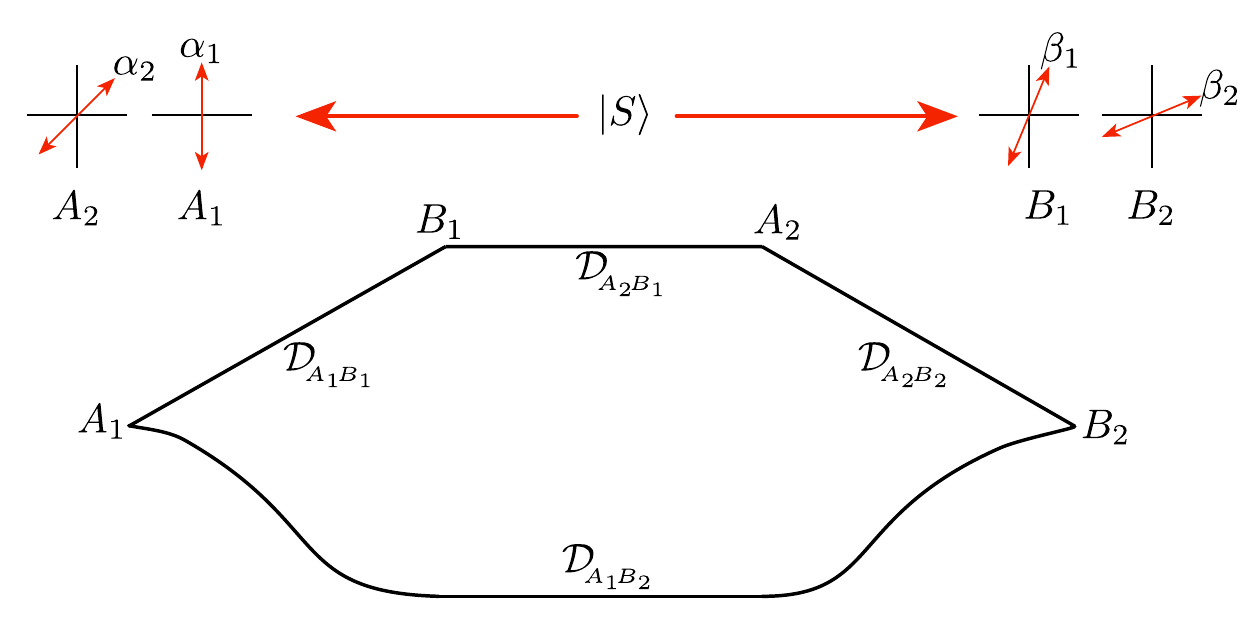}
   \end{tabular}
   \end{center}
   \caption{\em We illustrate here the information geometry of a Bell state analyzed by Schumacher \cite{Schumacher:1991}. There are two observers, Alice and Bob, that are detecting the 2-photons from the Bell state $|\Phi^+\rangle$. Alice has two detectors, one linear polarizer rotated an angle $\alpha_1$ away from vertical, the other detector is rotated an angle $\alpha_2$, and similarly for Bob. An ensemble of Bell states is prepared and Alice and Bob randomly choose one or the other detector. This leads after many measurements to four binary random variables, $A_1$, $A_2$, $B_1$ and $B_2$. At the bottom of the figure, we show the quadrilateral formed by these for random variables. We can use Eq.~\ref{eq:length} to calculate the four distances ${\cal D}$'s shown on the edges. We cannot connect the diagonals as they are mutually exclusive; therefore, we can not define an information area.
 \label{fig:S} 
   }
   \end{figure}

The Shannon--based information distance in Eq.~\ref{eq:length} shares the same bounds for any classical or quantum bipartite state. Then how can it be used to capture the stronger--than classical correlations? The answer: Schumacher showed that one must utilize the superposition principle that gave way to multiple mutually unbiased bases ($MUB$). In particular, he showed that at least two different detectors needed to be used for each of the two photons in the Bell state. These detectors can reveal a relationship between the entanglement and its information geometry. Based on this approach, Schumacher examined the relationship between the violation of the Bell inequality for a singlet state and the triangle inequality in information geometry \cite{Schumacher:1991}. This is illustrated in Fig.~\ref{fig:S}. Here we briefly review his results.

First, we consider two observers, Alice ($A$) and Bob ($B$). We provide many identical copies of the Bell state to them as shown in Fig.~\ref{fig:S}. Alice receives the photon propagating to the left, and Bob receives the photon traveling to the right. Alice chooses randomly one of two detectors. Alice's first detector, $\langle \alpha_1 |$, is a linear polarizer rotated clockwise from the vertical state $|\updownarrow\rangle$ by an angle $\alpha_1$, and her second detector is rotated by an angle $\alpha_2$. Similarly, Bob's first and second detectors are rotated by $\beta_1$ and $\beta_2$; respectively.
Schumacher parameterized the four angles,   
\begin{equation}
\label{eq:angles}
\left\{ \alpha_1=0, \alpha_2=2\theta, \beta_1=\theta,  \beta_2=3\theta
\right\}, 
\end{equation}
using  a single angle $\theta=\pi/8$.
It is important to clarify that we are using the usual Stokes-Poincare angles ($\alpha_i$'s) in this manuscript.  The physical angular settings of the polarizers are a factor of 2 smaller than the angles in Eq.~\ref{eq:angles}. Furthermore, the setting on the half-wave plate (HWP) is another factor of two smaller than the physical polarization angles, i. e. a factor of 4 smaller than the four Stokes angles.

If the quadrilateral formed by the four detectors as illustrated in Fig.~\ref{fig:S} was embedded in a Euclidean surface, then the direct route $A_1 \rightarrow B_2$ should always be less than or equal to the indirect route $A_1 \rightarrow B_1 \rightarrow A_2 \rightarrow B_2$,

 \begin{equation}
 \label{eq:tri}
 {\cal D}_{A_1B_2} \le  {\cal D}_{A_1B_1} +  {\cal D}_{A_2B_1} +  {\cal D}_{A_2B_2}.
 \end{equation}
However, Schumacher showed that this triangle inequality is violated for certain angles \cite{Schumacher:1991}. 
We define the degree of triangle violation by a difference of the distances in Eq.~\ref{eq:tri},  where
\begin{equation}
\label{eq:tv}
{\cal V}(\theta) :=   {\cal D}_{A_1B_2}-\left({\cal D}_{A_1B_1} +  {\cal D}_{A_2B_1} +  {\cal D}_{A_2B_2}\right).
 \end{equation}
 
Schumacher calculated that the 2-photon Bell state yields a violation for $\theta=\pi/8$. In this particular case where three of the pairwise detectors have the same difference in their relative angular settings, whilst the relative angular setting  between the direct connection between $A_1$ and $B_2$ is three times larger, 
 \begin{equation}
\beta_1-\alpha_1 = \beta_1=\alpha_2 = \beta_2-\alpha_2 =\theta,
\end{equation} 
and therefore,
\begin{equation}
\beta_2-\alpha_1 = 3 (\beta_1-\alpha_1)=3 \theta,
\end{equation}
yielded a violation in the triangle inequality,
\begin{equation}
{\cal V}(\pi/8) \approx 1.3832.
\end{equation}
Schumacher introduced the first information geometry realization of a Bell-type inequality for the maximally entangled Bell state $|\Phi^+\rangle$.  In Schumacher's case with $\theta =\pi/8$, he found
\begin{equation} \label{eq:ie}
\underbrace{{\cal D}_{A_1B_2}}_{1.7832}    \not\le  \underbrace{   
\underbrace{{\cal D}_{A_1B_1}}_{0.4667}  + 
\underbrace{{\cal D}_{A_2B_1}}_{0.4667} +  
\underbrace{{\cal D}_{A_2B_2}}_{0.4667}
}_{1.4000}.
\end{equation}
It is the purpose of this paper to demonstrate an experimental realization of this inequality (Eq.~\ref{eq:ie}) using SPDC of a photon by a paired set of Type-1 BBO crystals. We will demonstrate a statistically-significant triangle violation, ${\cal V}(\theta_i)>0$, over a range of eight different angles. We describe our experimental setup, our methods and our results confirming Schumacher's prediction for the singlet state in the next section. We plot the measured violations in Fig.~\ref{fig:tiv}. And for the case of Schumacher's angle $\theta=\pi/8$, we show our results in Eq.~\ref{eq:expie} in Sec.~\ref{sec:results} are in understandably good agreement with Eq.~\ref{eq:ie}. We show that our data illustrated in Fig~\ref{fig:tiv} is better represented by a modified Werner state, 
\begin{equation}
\label{eq:werner}
\rho_{\!{}_{MW}} = \lambda |\psi\rangle\langle \psi| + \frac{(1-\lambda)}{4} I\!\!I
\end{equation}
 with $|\psi\rangle\  \!\approx  |00\rangle+ e^{i 0.225}  |11\rangle$, entanglement parameter $\lambda \approx 0.998$ and $I\!\!I$ is the identity matrix. In the next section we describe our experiment. 

\section{\uppercase{Experimental Setup, Methods and Measurement Results}}
\label{sec:exp}

\begin{figure} [ht]
   \begin{center}
   \includegraphics[height=2.5in]{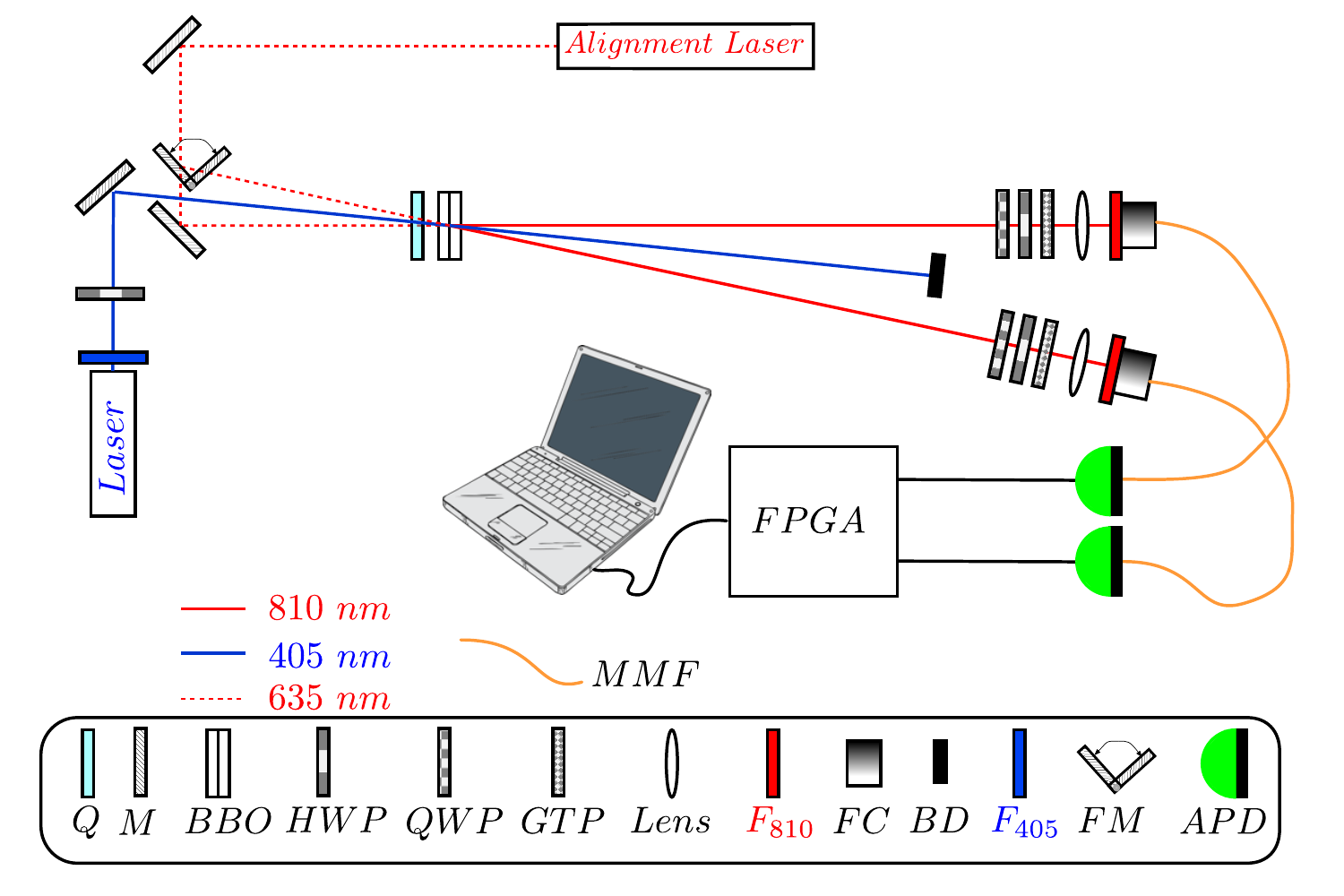}
   \end{center}
   \caption{{\em This figure illustrates the standard implementation of a Type--1 SPDC using a  paired BBO crystal used in this paper \cite{Galvez:2005,Kwiat:1999}.   The optical elements are as follows: $Q$, $0$'th--order quartz plate; $M$, mirror; $BBO$, paired Type--1 $\beta$--borate crystal;  $HWP$, half wave plate; $QWP$, quarter wave plate; $GTP$, Glan Thompson polarizer; $F_{810}$, notch filter at $810\,nm$; $FC$, fiber coupler; $BD$, beam dump; $F_{405}$, notch filter at $405\,nm$; $FM$, flipper mirror;  $APD$, single-photon detector avalanche photodiode; FPGA, field--programmable gate array; and $MMF$, multi--mode fiber. 
 \label{fig:setup} 
   }}
   \end{figure}

Our experimental implementation of SPDC is illustrated in Fig.~\ref{fig:setup} and is described by Kwiat \cite{Kwiat:1999}. The key elements of our setup are (1) the laser source, (2) the nonlinear medium, and (3) a single-photon detection system and coincidence counting. Each of these stages of our optical arrangement are as follows: 
\begin{itemize}

\item A Continuous-wave 405nm, 50mW blue diode laser beam is preconditioned by passing it through a narrow bandpass filter $F_{405}$, followed by a 405nm half--wave retarder and then reflecting it off of a mirror with vernier adjustments that directs the beam toward the non-linear crystal array. 
\item The crystal array consists of a quartz spatial compensation plate, and two paired Type-1 BBO crystals and a temporal compensation plate. This plate is a ``0"--order 5.5~mm quartz crystal. Each BBO crystal has a thickness of $5\,mm$.

\item Each of the two collection and detection systems for the correlated photon pairs consists of an $810 nm$ collimator and fiber coupling to a multimode fiber. Before coupling into the multimode fiber, the down--converted photons each pass through an optical polarimeter consisting of a QWP, and HWP and a Glan-Thompson polarizer. Just before the collimator, we place an $810 nm$ narrow--band filter centered on the wavelength of the SPDC photons. These collimators are arranged in a $1 m$ arc centered on the crystal in order to collect the two photons on opposite sides of the $\sim \!6^{\circ}$ cone. Each of the two identical multimode fibers is coupled into a single-photon detector, an Excelitas Technologies SPCM-AQRH-13-FC with a dark count $<250\ CPS$. The coincidence counting was done on the outputs of the detectors by a field-programmable gate array (FPGA), an Altera DE2 Board P0301. The time bins we used were $40\ ns$. The programming of the FPGA unit follows the approach of Branning \cite{Branning:2009,Galvez:2005}.
\end{itemize}

\subsection{Characterization and quality of our SPDC Quantum State}
\label{sec:qsc} 
We use the usual measures to fully characterize and determine the quality of the quantum states produced by our Type-1 paired BBO SPDC optical setup that is illustrated in Fig.~\ref{fig:setup}.   In particular we report on (1) the $H$-$V$ and $A$-$D$ visibility, (2)  QST and its fidelity, (3) the Clauser, Horne, Shimony, and Holt (CHSH) inequality, as well as (4) the tangle and concurrence.   Our collected polarimetry data consisting of the usual spectrum of 16 different coincidence measurements, 
\begin{eqnarray}
\label{eq:16}
{\mathcal M}_{tomo} &=& \left\{
HH,HV,VV,VH,RH,RV,DV,DH,DR,DD,\right.\\
& &  \left. RD,HD,VD,VL,HL,RL
\right\}.
\end{eqnarray}
We followed the standard procedures as reported in James et al. \cite{James:2001}.
The QST was constructed from coincidence counts and from the maximum likelihood estimation technique. 
The density matrix that we obtained  directly from the 16 coincidence measurements yields a $4\times 4$ matrix of complex numbers 
\begin{equation}
\label{TOMO}
{\tiny
\rho_S = 
\left(
\begin{array}{cccc}
0.493 & -0.031+0.017\,i & 0.005-0.003\,i & 0.506+0.295\,i\\
-0.031-0.017\,i  & 0.005 & 0.029-0.217\,i& 0.011+0.013\,i\\
 0.005+0.003\,i& 0.029+0.217\,i& 0.0053 & -0.019-0.049\,i\\
 0.506-0.295\,i& 0.011-0.013\,i &  -0.019+0.049\,i & 0.495\\
\end{array}
\right).
}
\end{equation}
Each of the $16$ modes in Eq.~\ref{eq:16} consists of $50$ coincidence measurements.  These are collected with a time resolution of $40\,ns$ as  determined by FPGA.  The photon counts in each time bin for idler beam ($A$) was, on average, 9900, and for the signal beam ($B$) 7500. For example, the average coincidence count for $H\!H$ was 1167, and $HV$ was 19 (these include the accidental counts of $\sim 6$ counts).
We find for our SPDC state produces a quantum state closely approximating the Bell state with fidelity ${\mathcal F}= 0.910$, tangle ${\mathcal T} = 0.906$ and linear entropy,  ${\mathcal H}=0.086$ and the concurrence, ${\mathcal C}= 0.951$, here ${\mathcal C}= 1$  for the maximally entangled Bell state ${\mathcal C}= 0$ for the totally mixed state.

The tomography measurement  is already an indication that we have a good quality state approximating the Bell state. However, we also measured the CHSH inequality for our SPDC photons. Theoretically, the CHSH inequality for the Bell state is the quantum upper bound
\begin{equation}
{\mathcal M}=E(A_1,B_2) - \left(E(A_1,B_1) +E(A_1,B_1) +E(A_1,B_1)\right) \le 2 \sqrt{2} \approx 2.828,
\end{equation} 
where $E(A,B)$ are the quantum correlations \cite{CHSH}.
For a specific sets of measurements angles (i. e. the HWP angles)
\begin{equation}
\theta_{CHSH}=\left\{0,\frac{\pi}{8},\frac{\pi}{4},\frac{3 \pi}{8},\frac{\pi}{2},\frac{5 \pi}{8},\frac{3 \pi}{4},\frac{7 \pi}{8}\right\}
\end{equation}
our measurement of the CHSH inequality is
\begin{equation} 
{\mathcal M}=2.735\pm0.003.
\end{equation}
This result is within  $1.73\%$  error with respect to expected Bell state theoretical value \cite{Weihs:1998}. 

Perhaps most importantly, we measured our quantum states horizontal--vertical visibility, $V_{hv}= 0.998$ and the diagonal--anti-diagonal visibility $V_{ad}=0.970$ \cite{Galvez:2005}.

\subsection{Measured Results for Schumacher's Triangle Inequality}
\label{sec:results}

We conducted nine distinct measurements for the Schumacher triangle inequality violation.  One of our measurements was at the point of maximal triangle inequality violation at $\theta=0.328$ radians, The other eight measurements straddled the maximum.  The nine Stokes angles (measured in radians) are
\begin{equation} 
\left\{ \theta_i\right\}_{i=1,2,\ldots 8} = \left\{ 0.175,0.227,0.279,0.328,0.393,0.436,0.471,0.503\right\}.
\end{equation}

Each of the nine angle settings, $\theta_i$, involved 16 different settings of the HWPs of $A$ and $B$. This corresponds to four settings for each of the four edges of the quadrilateral in Fig.~\ref{fig:S}. We refer to each of these settings as modes. For each of the 16 modes, we measured over $350$ coincidence measurements. This enabled us to generate the joint and individual probability distributions for Alice and Bob for each angle $\theta_i$ and for each of the four edges of the quadrilateral. We used these probability distributions to determine the entropies using Eqs.~\ref{eq:ha}--\ref{eq:hab}. We then could calculate the four information distances ${\cal D}_{A_1B_1}$, ${\cal D}_{A_2B_1}$,${\cal D}_{A_2B_2}$ and then the base ${\cal D}_{A_1B_2}$ for each of the angular settings, $\theta_i$.

We identified three major sources of error, (1) the state preparation error that is a systematic error, (2) calibration of the polarimetric elements that is also a systematic error, and (3) the statistical error in measuring the coincidences. We subtract the accidental counts for each of our measurements.  We propagated the error using the standard errors in the coincidences, $N_i\pm \delta N_i$, and the errors in the calibration for the angles of $A$, where $\alpha_k\pm\delta \alpha_k$, and $B$, where $\beta_\ell\pm\delta \beta_\ell$ , in the usual way adding the statistical uncertainties in quadrature
\begin{equation}
\delta {\cal D}_{A_iB_j} = \sqrt{\left(\frac{\partial {\cal D}_{A_iB_j}}{\partial \alpha_k} \delta \alpha_k\right)^2 + 
\left(\frac{\partial {\cal D}_{A_iB_j}}{\partial \beta_\ell} \delta \beta_\ell\right)^2 + \sum_{j=1}^4 \left(\frac{{\partial \cal D}_{A_iB_j}}{\partial N_j} \delta N_j \right)^2 }.
\end{equation}
Our calibration was done using four people each making 10 measurements on each element of the polarimeter.  The HWP and the QWP were placed between our crossed Glan--Thompson polarizers and minima were found. We then calculated the standard deviation of our 40 measurements. We assigned the maximum calibration error to our two HWPs in our error analysis, $\delta \alpha_k=\delta \beta_\ell=\delta \phi_{max}\approx 0.0030$.  We used motorized mounts for each HWP.  

\begin{figure}[htbp] 
   \centering
   \includegraphics[width=5in]{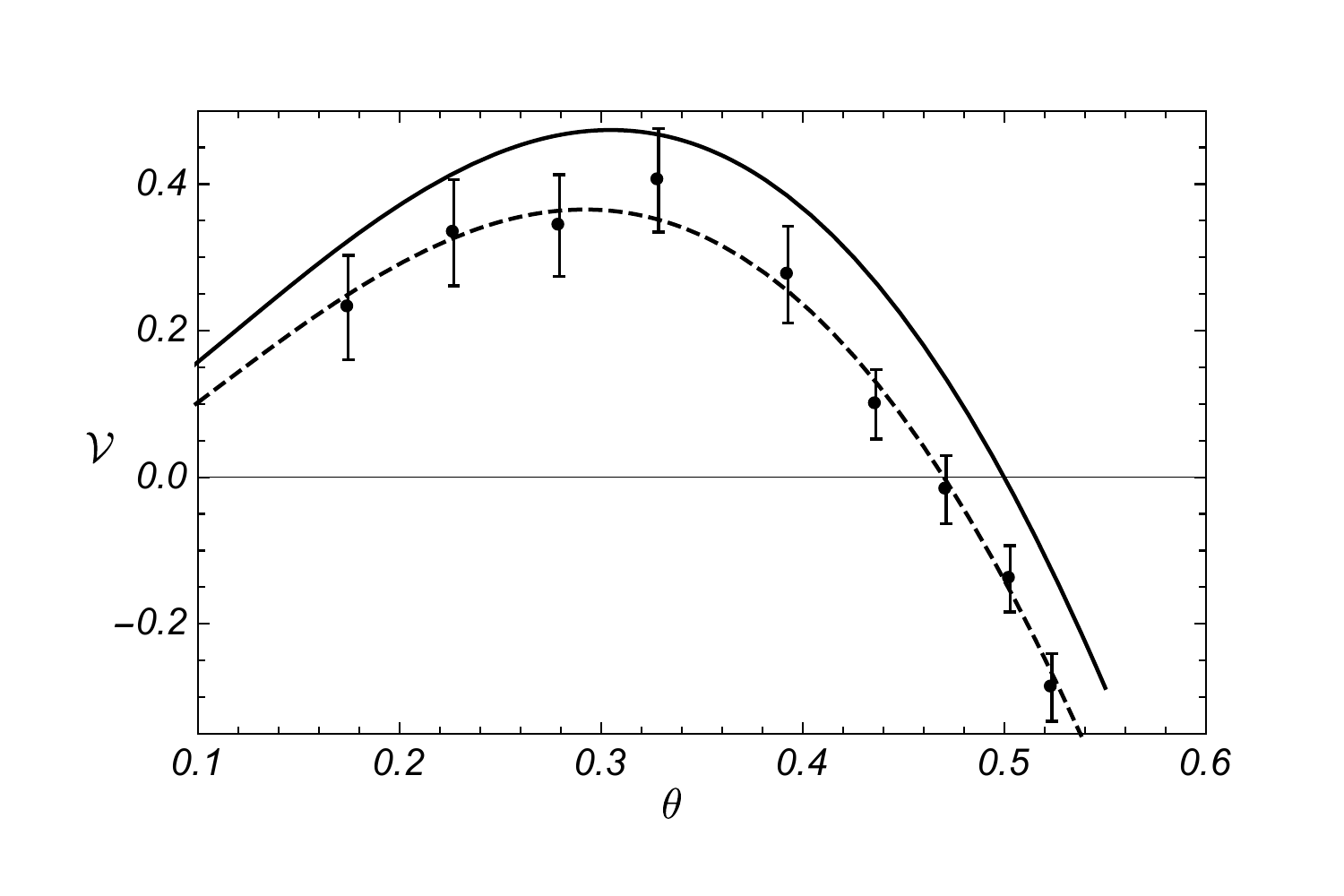} 
   \caption{  {\em The experimental realization of the triangle inequality violation of Schumacher. We plot the degree of triangle violation, ${\cal V}$ defined in Eq.~\ref{eq:tv} as a function of the nine angular parameters for Alice and Bob's detector settings,  $\theta$ of Eq.~\ref{eq:angles} in radians.  Positive values of ${\cal V}$ signal triangle inequality violation. The solid curve represents the theoretically expected results based on Schumacher for a pure Bell state.  The dashed line is the results that fit our data for the Werner state in Eq.~\ref{eq:werner} with $\lambda \approx 0.998.$}
   }
   \label{fig:tiv}
\end{figure}

Our results for the nine measurements of the Schumacher information geometry triangle inequality are displayed in Fig.~\ref{fig:tiv} and  where we show a statistically-significant agreement with his entropic geometry model. Our measured triangle inequality measurements systematically fall below the theoretical curve for the pure Bell state that is predicted by Schumacher.  However, we expect that the state we generate from SPDC will differ slightly from this pure state, as we observed in the QST, the visibility and the CSHS measurement.  To better fit the observed data we expect that there is a small degree of decoherence introduced in our optical arrangement.  A maximally austere and least unreasonable assumption we can use to fit the data would be to add to the singlet state a fraction of a totally mixed state. Since Werner state is a mixture of a Bell state and maximally--mixed state, we thought it would be reasonable to assume our state is Werner state. This assumption does match our result of quantum tomography and density matrix  \ref{TOMO}.  Following this assumption, we found that the modified Werner state,
\begin{equation}
\label{eq:wmm}
\rho  = \lambda |\psi\rangle\langle \psi| + \frac{(1-\lambda)}{16} I\!\!I,
\end{equation}
with 
\begin{equation}
\label{eq:psi}
 |\psi\rangle\  =  |00\rangle+ e^{i 0.225}  |11\rangle 
\end{equation}
 and $\lambda\approx0.998$ given in Eq.~\ref{eq:wmm} gives a better least squares fit to our data as illustrated in Fig.~\ref{fig:tiv}.  The least-squared fit gave a reasonable relative phase, $\phi\approx 0.225$, and small fraction, $0.2\%$ of diagonal density matrix.  The good qualitative fit seems to us to be a reasonable model, though far from an exhaustive fit in higher dimensions.  Furthermore, the value of $\lambda$ is close to the concurrence of the modified Werner state, which is equal to $0.997$. The CHSH equality calculated for this state is equal to 2.360.

The maximum triangle inequality violation occurs at $\theta_4=0.0328$, this is also the maximal measurement we observed.   If we focus more closely on the angle that Schumacher used in his calculation, $\theta_5=0.393$ we can compair our four distances with that of the theory.  We found the following four distances of the edges of the Schumacher quadrilateral  illustrated in Fig.~\ref{fig:S}:
\begin{equation} \label{eq:expie}
\boxed{
\underbrace{{\cal D}_{A_1B_2}}_{1.733\pm 0.124}    \not\le  \underbrace{   
\underbrace{{\cal D}_{A_1B_1}}_{0.511\pm 0.041}  + 
\underbrace{{\cal D}_{A_2B_1}}_{0.512\pm 0.041} +  
\underbrace{{\cal D}_{A_2B_2}}_{0.533\pm 0.041}
}_{1.556\pm0.071}.
}
\end{equation}
This is consistent with Schumacher's triangle equality violation for the Bell state in Eq.~\ref{eq:ie} in light of the Werner state model.

\section{\uppercase{Entanglement Geometry: Future Directions}}
\label{sec:O}
We experimentally demonstrated that Schumacher's Shannon-based information geometry triangle inequality is violated for a Bell state \cite{Schumacher:1991}. We explicitly showed that in the landscape of quantum entanglement, the direct distance, in the information sense, is not always the shortest distance. This is a common experience for anyone with experience commuting in a congested city. 

While Schumacher's theoretical construction is analogous to the well--known result of Bell, it presents itself as a novel information theoretic way of identifying, and even quantifying quantum correlation or quantum entanglement. Schumacher found a geometric configuration that is based on measurement outcomes and classical information geometry that is sensitive to quantum correlations. This provides a novel view into quantum entanglement using Shannon's entropy over the SOM.  Naively, one might think this is impossible for Shannon's entropy to distinguish classical correlations from quantum correlations; after all, they both share the same bounds on entropy. The key to overcoming this obstacle was to utilize the superposition principle, recognize that there are multiple MUB spanning the Hilbert space, and construct a measurement geometry where the observers (Alice and Bob) of the Bell state each have two separate non-orthogonal detectors at their disposal. Building up a recording for coincidence measurements for each of the four detectors ($A_1$, $A_2$, $B_1$ and $B_2$) from an ensemble of identically--prepared quantum states, gives rise to four pairwise binary random variables, the graph of which is a quadrilateral. Schumacher predicted that, for a range of detector settings, that the singlet--state would not be embeddable in the Euclidean plane. We verified this experimentally.
 
 \begin{figure}[htbp] 
   \centering
   \includegraphics[width=6in]{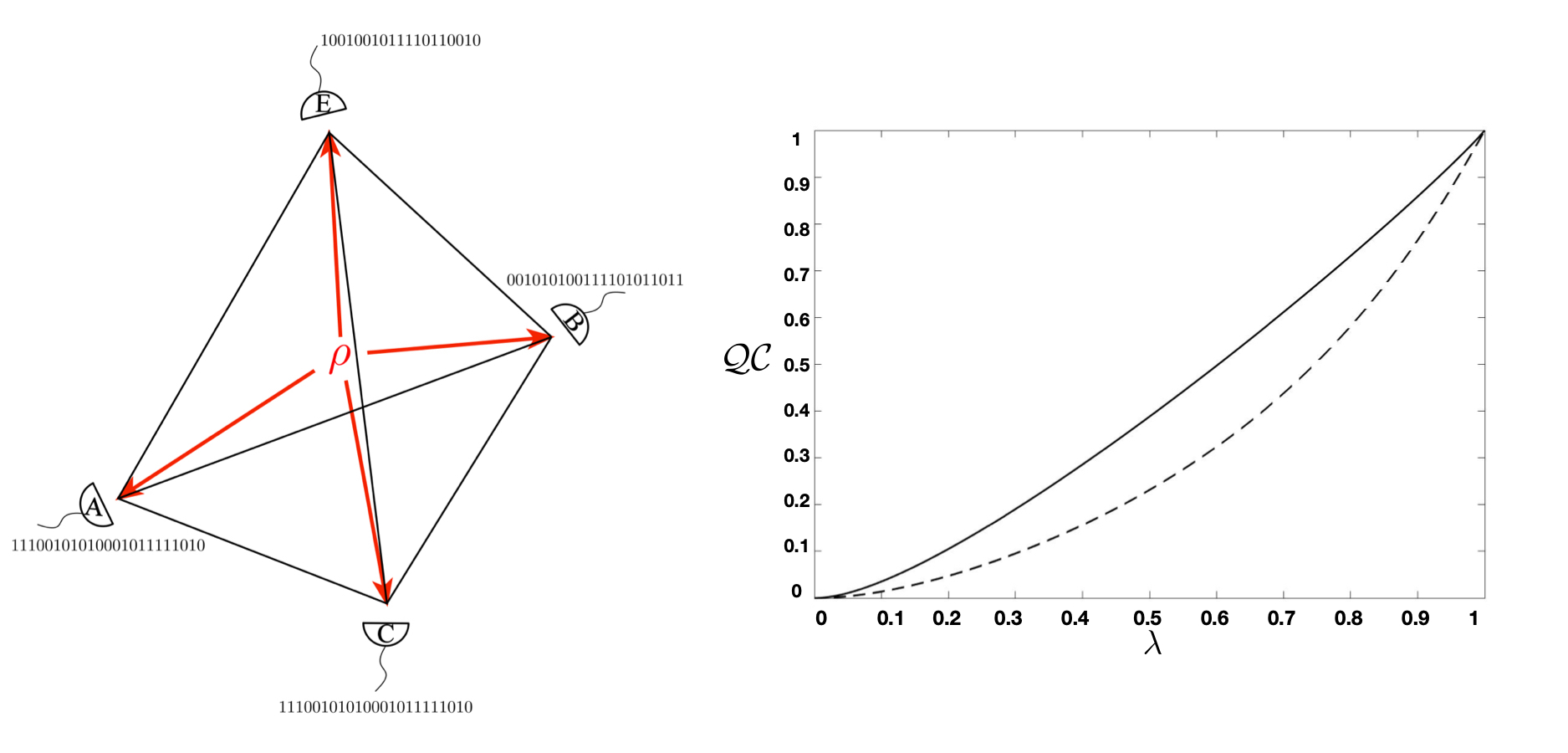} 
   \caption{  {\em The measurement space of Alice, Bob, Charlie and Eve for a four-qubit  density matrix Eq.~\ref{eq:w4} is displayed to the left of the graph. Each observer has a  has a detector labeled $A$, $B$, "C" and $E$; respectively. For a given setting of their detector they record the binary outcome results (shown as the string of ``$1$''s and ``$0$''s ) of the measurement over an ensemble of equally prepared Werner states. They form a tetrahedron $\overline{ABCE}$.   The reactivity is a ratio of the tetrahedral volume divided by the four triangle faces averaged over all possible detector settings. We display a plot of the quantum correlation, $\mathcal{QC}$ as measured by the global quantum discord (solid line)  and the quantum reactivity (dashed line) for the 4-qubit Werner state as a function of entanglement parameter $\lambda$.  Both measures yield a monotonically increasing function in lambda.}
   }
   \label{fig:werner4}
\end{figure} 
 
What is interesting to us is that there appears to be a clear connection between geometry and topology and the quantum  entanglement.  This is the case with Schumacher's triangle inequality.  We believe his approach can be generalized from bipartite states to higher--dimensional multipartite quantum states, as illustrated in Fig.~\ref{fig:werner411} for a four-qubit Werner state.
\begin{equation}
\label{eq:w4}
\rho  = \lambda |\Phi^{+}\rangle\langle \Phi^{+}| + \frac{(1-\lambda)}{16} I\!\!I
\end{equation}
with entanglement parameter $\lambda \in [0,1]$.
In first step, we generalized the information distance to area and higher dimensional  volumes.  This generalization is explained in detail in \cite{Miller:2019}.  Second, we used these areas and volumes and examined at the geometrical properties of quantum networks.  We showed that quantum correlation has a relationship with a geometrical quantity which we call quantum reactivity.  Quantum reactivity is in the usual sense ratio of surface area to volume \cite{Aslmarand:2019,Miller:2018}. 
\begin{equation}
\label{eq:reactivity}
 \overline{\mathcal R} :=  \overline{\mathcal A} / \overline{\mathcal V}.
 \end{equation}
=Here,  $\mathcal A$ is information area and $\mathcal V$ is volume that are generalizations of the information length in Eq.~\ref{eq:length} and were defined for four random variables, $A$, $B$, $C$ and $E$ in \cite{Miller:2019} and illustrated in Fig.~\ref{fig:werner4}.  In particular, 
\begin{eqnarray}
{\mathcal A}_{ABC} &= H_{A|BC}H_{B|CA}+H_{B|CA}H_{C|AB}+H_{C|AB}H_{A|BC},\\
{\mathcal V}_{ABCE}& := H_{A|BCE}H_{B|CEA}H_{C|EAB} +H_{B|CEA}H_{C|EAB}H_{E|ABC}\\
&\ \ +H_{C|EAB}H_{E|ABC}H_{A|CEB}+H_{E|ABC}H_{A|BCE}H_{B|CEA}.
\end{eqnarray}
 Additionally, to make the quantum reactivity observer-independent, we average over all possible measurements as indicated by the bar atop of the area and volume in Eq.~\ref{eq:reactivity}. This averaging also guarantees that this measure is invariant under unitary transformations as shown in \cite{Aslmarand:2019b}.  Although quantum reactivity scalable to higher dimensional multipartite states, it is computationally and experimentally formidable.  However, we know that for certain classes of quantum states the fidelity of the measure of quantum correlation may scale favorably in the number of measurements \cite{Vedral:1997,Aaronson:2018}, and perhaps the fidelity of the quantum reactivity under partial measurements will converge to the reactivity exponentially fast in the number of measurements, e.g. in the sense of quantum Sanov's theorem.  It's also worth mentioning that quantum reactivity is sensitive to quantum correlation and not just entanglement; therefore, it can have a  wider range of application in quantum mechanics since its shown that quantum correlation is essential for quantum computation \cite{Vedral:2010,Jozsa:2003}.  This brings us beyond the scope of this manuscript, as does the applications of these information geometry constructs to the recent  work in gauge/gravity duality spawned by Maldacena \cite{Maldacena:1998,Raamsdonk:2010,Qi:2018,Han:2019}.   Perhaps these tools can be applied to complex quantum networks in meaningful ways.

\section*{\uppercase{Acknowledgements}}
 
This research benefited form the mentorship of Enrique (Kiko) Galvez, without which we would not have been able to make these measurements.  We wish to acknowledge stimulating discussions with Verinder Rana and  C. Harvey. We thank Dongxue Qu for her help with programming and setting up the coincidence counting hardware and software, Daniel Carvalho for his help checking the calculation, and our machinist Mark Royer for helping us with our lab. One of us (WAM) wishes to thank AFRL/RITQ and the Griffiss Institute  for providing a stimulating research environment and support under the Visiting Faculty Research Program. PMA and WAM would like thank support from the Air Force Office of Scientific Research (AFOSR).  This research was supported under AFOSF/AOARD grant \#FA2386-17-1-4070 with a  supplement from AFRL/RITQ, and from an AFOSR/DURIP grant \#FA9550-19-1-0389.  D. A. is partially supported by grants from  IITP 2017-0-00266, NRF grant No. NRF-2020M3H3A1105796  and AFOSR grant FA2386-21-1-0089. We also acknowledge partial support from internal FAU funds from the Center for Connected Assured Autonomy (C2A2) and the Charles E. Schmidt College of Science.    Any opinions, findings, conclusions or recommendations expressed in this material are those of the author(s) and do not necessarily reflect the views of AFRL.    

\bibliography{qig2019} 

\end{document}